\documentstyle[12pt]{article}

\begin{document}
\title{
Universal bifurcation property of two- or higher-dimensional
dissipative systems  in parameter space:
Why does 1D symbolic dynamics work so well?}
\date{}
\author{
H. P. Fang
\\{\small CCAST (World Laboratory) P. O. Box. 8730,
        Beijing, 100080, China;}
        \\{\small Department of Physics, Fudan University,
        Shanghai, 200433, China.$^*$}
        \\{\small Institute of Theoretical Physics, P.O. Box. 2735,
        Beijing, 100080, China.}
	}
\vskip.15in
\maketitle
\begin{abstract}

The universal bifurcation property of the H\'enon map in parameter
space is studied with symbolic dynamics. The universal-$L$ region
is defined to
characterize the bifurcation universality. It is found that the
universal-$L$ region for relative small $L$ is not restricted to
very small $b$ values. These results show that it is also a universal
phenomenon that universal sequences with short period can be found
 in many nonlinear dissipative systems.

\end{abstract}

\vskip.15in

\section*{\center\small\bf I. INTRODUCTION}

 One of the standard ways of investigating the dynamics of physical
systems is by exploiting the universal (system-independent)
property$^{1-8}$ of them.  The best understood transition sequence
is the period-doubling cascade, which has been observed in a variety
of physical systems. Beyond the accumulation point for the
period-doubling sequence there is chaos. Two decades ago
Metropolis, Stein and Stein showed$^{1}$ that there is an ordered
 sequence of distinct periodic windows, each of which occurs for
some range of control
parameter, within the chaotic region for unimodal maps,
$x_{n+1} = f(\mu, x_n)$. They have called this sequence the U-sequence since
the ordering of the windows is system independent. Remarkably, this
universality is also observed in systems with many degrees of freedom
both experimentally$^{2,3,8}$ and theoretically$^{4-7}$ although the
phase portraits of these two- or high-dimensional system still
exhibit very complex behaviour which is clearly not one-dimensional
or close to one-dimensional. It has been found that the
periodic windows interspersed in chaotic region for these systems are
ordered in a systematic way as those of one-dimensional (1D) maps.
The most striking and detailed observation is obtained in the
Lorenz equations
\begin{equation}
\begin{array}{l}
\dot x = 10 ( y - x ), \\
\dot y = r x - x z - y ,\\
\dot z = x y - 8 z/3. \\
\end{array}
\end{equation}
On the parameter $r$ axis with $45<r<400$, all of the found 68
periodic windows of the Lorenz equations can fit into those of
 a 1D antisymmetrical map with only one exception$^9$.
 Experimentally, even though the Belousov-Zhabotinskii reaction
involves more than thirty chemical species, it exhibits rather
complex bifurcation behaviour that  is modeled well by 1D maps$^3$.

Despite these numerical and experimental observations, the underlying
mechanism for the universal property is not fully understood. The
motivation of this paper is to present an approach towards
interpreting all these experimental and numerical observations and
exploring their limitations.
We will take the H\'enon map $^{10}$
\begin{equation}
\cases{x_{n+1} = 1 - a x_n^2 + y_n, \cr y_{n+1} = b x_n,}
\end{equation}
as an example. The bifurcation structure of the H\'enon map in the
two-dimensional parameter $(a,b)$ space has been extensively
discussed$^{11,12}$. In this paper, we will use symbolic
dynamics$^{1,14-20}$ of 1D mappings and 2D mappings to illustrate
the universal topological property of the H\'enon map at selected
parameters by considering the unstable periodic orbits embedded in
its chaotic attractor. Two topological quantities $\delta$ and
$L$ are defined to characterize this universal topological property.
 Then we discuss the universal bifurcation property of the H\'enon
map in 2D parameter space  $(a, b)$ by defining universal-$L$
regions in which the H\'enon map exhibits 1D bifurcation behavior
to period $L$. It is remarkable to find that the universal-$L$ region
for relative small $L$ is not restricted to very small $b$ values.
We will also present two examples of ordinary differential equations
(ODE's), the R\"ossler equations$^{13}$ and the forced
Brusselator$^4$, to demonstrate the validity and robustness of our
approach.  These results show that it is also a universal
phenomenon that universal sequences with short period can be found
in many experiments or numerical calculations on nonlinear
dissipative systems.

The paper is organized as follows. In Section II, we review the basic
property of 1D unimodal maps. The universal bifurcation property and
its limitations of the H\'enon map in the 2D parameter $(a,b)$ space
 is studied in Section III. To demonstrate the validity of the method
 presented in Section III, the universal bifurcation property of
 R\"ossler equations and the
 forced Brusselator in a definite parameter axis is investigated in
Section IV. Finally, in Section V we give  our conclusion.

\section*{\center\small\bf II. UNIVERSAL SEQUENCES IN 1D UNIMODAL
MAPS}

By using the symbolic dynamics of 1D mappings, Metropolis, Stein and Stein
(MSS) had already shown that the dynamics of unimodal 1D maps of the
interval [-1, 1] is embodies in the U-sequence of periodic
windows$^{1, 14, 15}$.  Fig.~1 shows a typical case. The extremum is
denoted by a letter C. Each periodic window of the map can be labelled by
a symbolic sequence of 0's and 1's that mark the location (to the left or
right of C ) of the successive iterates of the initial point C. For
example, the only windows with period 5 are  101$^2$C, 10$^2$1C, and
10$^3$C ( 101$^2$C represents the periodic window (101$^2$C)$^\infty$
hereafter.) Indeed, we can define an ordering$^{14, 15}$ for these
symbolic sequences referring to the natural order in the interval
[-1, 1]. These ordering rules are consistent with the ordering of a
real number $\alpha$ defined for a sequence S($x$) with an initial
point $x$ as following$^{17}$
\begin{equation}
\alpha ({\rm S}(x)) = \sum_{i=1}^\infty \mu_i 2^{-i},  \ \ \ \ \
\end{equation}
with [16]
$$\mu_i = \cases{0 \cr 1}  \ \ \
{\rm for} \   \cases{ \sum_{j=1,}^i s_j = 0   \cr
\sum_{j=1,}^i s_j = 1} \ \ \ {\rm (mod} \ 2) .$$
Since the symbolic sequence K=S(C), called also the kneading sequence,
acquires
a maximal $\alpha$ in this metric representation, a symbolic sequence
S(x) corresponds to a real trajectory if and only if it satisfies
\begin{equation}
\alpha(\sigma^m ({\rm S}(x))) \leq \alpha ({\rm K}), \ \ \
m = 0, 1, 2, \cdots,
\end{equation}
where $\sigma$ denotes the shift operator. With this admissibility condition,
we can generate all the admissible periodic orbits for a given kneading
sequence K. The kneading sequence changes as the controlling parameter
alters. Since kneading sequences correspond to orbits coming from C, they
should also satisfy the above condition. Thus we obtain the admissibility
condition for K  themselves: a symbolic sequence K can be a kneading
sequence if and only if it satisfies
\begin{equation}
\alpha(\sigma^m ({\rm K})) \leq \alpha ({\rm K}), \ \ \
 m = 0, 1, 2, \cdots.
\end{equation}
When K is a periodic string, K corresponds to a periodic window. From Eq.
5, we can generate all the possible  periodic windows. It can be
checked that there are only three period 5 windows as those listed above.

With the ordering rules in equation (3), all periodic windows can  be
ordered to yield the U-sequence. In the logistic map, this U-sequence is
consistent with the increasing $\mu$ order which is listed in Table~1
up to period 7.

\begin{footnotesize}
\begin{table}[ht]
\caption[]{Symbolic sequences for periodic windows of the H\'enon map along
two different parameter axes and that for the forced Brusselator equations.
 The axis I (long dashed in Fig.~4) and axis II (dash-dotted) are two axes
in and out of the universal-7 region of the  H\'enon map, on which a
complete and incomplete U-sequence is found respectively.}
\begin{center}
\begin{tabular}{|l|l|l|lll|l|}
\hline
\hline
{No.} & Period & Word &  \multicolumn{3}{c}{$a$ range}
\vrule&  Brusselator$^4$ \\
\cline{4-6}
 & &  & b=0 & Axis I & Axis II & \ \ \ \ \ $\omega$  \\
\hline
 0 &  0  & C &   0-0.749 & 1.032565-1.164538 & none  & \\
   &  2  & 1C &   0.75-1.249 & 1.164539-1.332034  & none & 0.45-0.544\\
   &  2$\times$2  & 101C  &   1.25-1.367  &1.332035-1.409865
& none & 0.545-0.5777\\
 2 &  6  & 10111C &  1.4747-1.47973 &1.499840-1.495505
& none & 0.58249-0.58251\\
 3 &  7  & 101111C &   1.57472-1.57541 &1.585860-1.585225
& 1.521515-1.522540 &  \\
 4 &  5  & 1011C &  1.6244-1.62843 & 1.635395-1.631640
& 1.588650-1.594165 &0.5845-0.5848 \\
 5 &  7  & 101101C &  1.67396-1.6744 & 1.679020-1.678610
& 1.651665-1.652290 &  \\
 6 &  3  & 10C &  1.75-1.76853  & 1.763210-1.743840
& 1.776860-1.792900 & 0.5947-0.654 \\
   &  2$\times$3  & 10010C &  1.76854-1.77722
& 1.772255-1.763215 & 1.792920-1.800525 & 0.6545-0.7025\\
 7 &  7  & 100101C &  1.8323-1.83239  & 1.835474-1.835567
 &1.811364-1.811581  &  \\
 8 &  5  & 1001C &   1.86059-1.86136  & 1.863083-1.863841
& 1.843810-1.844705 & 0.7068-0.7115\\
 9 &  7  & 100111C &  1.8848-1.88483 & 1.886869-1.886911
&1.870425-1.870470 &   \\
 01 &  6  & 10011C &  1.90726-1.90736 & 1.909005-1.909120
&1.894790-1.894915 & 0.718-0.7185 \\
 00 &  7  & 100110C &  1.92715-1.92716 & 1.925261-1.925289
& 1.939390-1.939415\\
 12 &  4  & 100C &  1.94056-1.94153 & 1.938829-1.939827
& 1.952015-1.952945  &  0.7345-0.792 \\
 13 &  7  & 100010C &  1.95371-1.95371 & 1.952122-1.952138
& 1.964545-1.964555 & \\
 14 &  6  & 10001C &  1.96677-1.9668 & 1.968064-1.968099
& 1.956760-1.956795 & none \\
 15 &  7  & 100011C &  1.977179-1.977184 & 1.978379-1.978385
&1.967795-1.967795 & \\
 16 &  5  & 1000C &  1.98541-1.985468 & 1.984101-1.984160
& none & 0.8259-0.8675 \\
 17 &  7  & 100001C &  1.991814-1.991818 & 1.992925-1.992928
& 1.982968-1.982969 \\
 18 &  6  & 10000C &  1.996375-1.996379 & 1.995150-1.995153
& none & 0.9015-0.923 \\
 19 &  7  & 100000C &  1.999096 & 1.997890-1.997891 & none &\\
\hline
\end{tabular}
\end{center}
\end{table}
\end{footnotesize}

\section*{\center\small\bf III. UNIVERSAL SEQUENCES IN 2D H\'ENON MAPS}
 The H\'enon map (2)
has be extensively studied by using symbolic dynamics$^{16-19}$.
The set of all ``primary" tangencies between stable and unstable
manifolds determines a binary generating partition
which divides the attractor into two parts marked by letters  0 and 1.
Any trajectory is encoded by a bi-infinite string
S($x$) =  $\cdots s_{-m} \cdots s_{-1} s_0 \bullet
s_1 s_2 \cdots s_n \cdots,$
where $s_n$ denotes a letter for the $n$th image, $s_{-m}$ a letter for
the $m$th preimage, each is  either 0 or 1, the solid dot indicates the
``present" position.  In order to extend the grammar for unimodal
maps to this map, a ``backward" variable is defined as$^{17}$
\begin{equation}
\beta ({\rm S}(x))= \sum_{i=1}^\infty \nu_i 2^{-(i+1)},  \ \ \ \ \
\end{equation}
with
$$\nu_i = \cases{0 \cr 1}  \ \ \
{\rm for} \   \cases{ \sum_{j=0}^{-i} (1-s_j) = 0   \cr
 \sum_{j=0}^{-i} (1-s_j) = 1} \ \ \ {\rm (mod} \ 2) \ \ \
 {\rm for} \ b>0,$$
 $$\nu_i = \cases{0 \cr 1}  \ \ \
{\rm for} \   \cases{ \sum_{j=0}^{-i} s_j = 0   \cr
 \sum_{j=0}^{-i} s_j = 1} \ \ \ {\rm (mod} \ 2) \ \ \ {\rm for} \
 b<0,$$

 For this 2D map, each primary tangency C is associated with a
bi-infinite kneading
sequence K (with the first backward letter $s_0$ undetermined which may
be 0 or 1) and two symmetrical points ($\alpha$(K), $\beta_-$ (K)) and
($\alpha$(K), $\beta_+$ (K) = 1 - $\beta_-$ (K))  in the symbolic
 plane corresponding to $s_0$ = 0 and 1 respectively $^{18}$. Analogously
 to those in unimodal maps, for all admissible points ($\alpha$, $\beta$)
with $\beta \in [\beta_-$ (K), $\beta_+$ (K)], $\alpha$ should be less
than $\alpha$(K) and thus the pruning front$^{17}$ is obtained by cutting
out rectangles $\{ \alpha, \beta | \alpha > \alpha (K), \beta \in
[\beta_- (K), \beta_+ (K)]\}$ for all points on the partition. The union
of these rectangles gives fundamentally forbidden zone.  Consequently,
the grammar for a word admissible or forbidden in this map can be
expressed as: A bi-infinite word is admissible if and only if all its
shifts never fall into the fundamentally
forbidden zone$^{17,18}$. It is clear that there are infinitely many
kneading sequences (corresponding to infinitely many primary
tangencies) in a 2D map to determine the admissibility condition for a
word, while there is only one kneading sequence in a 1D map.

$Universality\ in\ the\ H$\'e$non\ map.$ Fig. 3 shows a typical symbolic
plane, $(a, b)$ = (1.4, 0.16). The corresponding attractor is shown in
Fig. 2 which has a rather complicated structure. Its fractal dimension
is 1.16$\pm$ 0.03.  Numerically 203 kneading sequences are found as
shown in Fig. 2. It is found that the minimal and maximal of all the
forward parts of these kneading sequences start with
K$_{min}$=101111010101 and K$_{max}$=101111011111 respectively,
corresponding to a minimal and maximal $\alpha$-values $\alpha_{min}$
= 0.837560 and $\alpha_{max}$ =0.838466 of all these kneading sequences.
We define two quantities $\delta$ and $L$ as
\begin{equation}
\delta = \alpha_{max} - \alpha_{min} = 0.000906,
\end{equation}
\begin{equation}
L = - [log_2 \delta] = 10,
\end{equation}
where $[log_2 \delta]$ denotes the integer part of $log_2 \delta$.
It is clear that $\delta$ =0 and $L \rightarrow + \infty$ in the 1D limit
 ($b$ =0). For ($a, b$) = (1.4,0.16), an unstable periodic orbit with
length $n\leq L=10$ can not tell the difference
between these kneading sequences. Indeed, no symbolic string with length
 $n \leq L$ lies in the interval between K$_{min}$ and K$_{max}$.
Thus for the unstable periodic orbits with length $n\leq L$, the grammar
is completely determined by a symbolic string K$_f$,
 which is 1011110101 or 1011110111, the first 10 letters of K$_{min}$ or
K$_{max}$, that is, a word S($x$) corresponds to an unstable periodic orbit
 of the H\'enon map for ($a, b)$ = (1.4,0.16) if and only if it satisfies
\begin{equation}
\alpha(\sigma^m (S(x))) \leq \alpha ({\rm K}_f), \ \ \  m =
0, 1, 2, \cdots.
\end{equation}
This is just the grammar for unimodal maps with a kneading sequence K$_f$.
Consequently the unstable periodic orbits  of the H\'enon map for
$(a, b)$ = (1.4, 0.16) can be generated as that of unimodal maps with a
 kneading sequence K$_f$ (see Eq. 4). The only exception is the unstable
  periodic orbit  $K_f^\infty$ which can not be determined by Eq. 9.
We here noted that the H\'enon map is divergent for $a=1.4$ and
$b>0.315$.

 As $b$ decreases, $L$ increases. It had already shown$^{19}$ that
$L$ = 32 for $(a, b)$ = (1.4, 0.05). In the 2D phase space, even for
$(a, b)$ = (1.4, 0.05) the attractor has a clear hook indicating that
the map is two-dimensional.  We emphasize that though the attractors
reveal very complicated structure in 2D phase space, the topologies
for these attractors may be very close to those in 1D maps that the
unstable periodic orbits can be generated with only one kneading
sequence to  some degree.

Now we consider the universal bifurcation property of the H\'enon map
in parameter space.  Fig. 4(a,b) show the isoperiodic lines$^{11,12}$
for all the nine period 7 windows.  Numerically  we find that $L \ \leq 7$
for all the parameters $a$ and $b$ in the
region between the two heavy solod lines shown in the figure$^{21}$.
We  call this region the $universal-7$ $region$ hereafter. Thus all the
 periodic orbits with length$\leq$7 of the H\'enon map in this Universal-7
region can be determined with only one kneading sequences as
those of 1D unimodal maps. Consequently, in this region there is a perfect
MSS-sequence up to period$\leq$7 along any axis provided that the axis
is never tangent with any isoperiodic lines.  These axes are in a sense the
 same  as the axis of $b$ = 0 (corresponding to the  Logistic map).
We present an example of these axes in Fig. 4 (line I, long dashed).
The periodic windows on this axis are listed in Table~1. It is clear that
they share the universal feature as that of 1D unimodal maps up to
period 7. We can also obtain $universal-M$ $region$ numerically  for
M = 5, 6, 8, 9, $\cdots$ in which  there are MSS-sequences for
period$\leq$ M along any axes provided that they  are never tangent with
any isoperiodic lines for period$\leq \ M$.

In Fig. 4 we also show the borders for the H\'enon map exhibit an
attracting set with initial points $(x_0, y_0)$ very close to original
point $(0, 0)$. Comparing to this borders we can say that
the universal-7 region is not restricted to very small $b$ values.
Thus it is rather likely to get a MSS-sequence up to a relative
short period (say, period 7) in the full 2D parameter plane of the
H\'enon map.

$Incomplete\ U-sequence\ in\ the\ H$\'e$non\ map.$ In fact, even on a axis
out of the universal region, the H\'enon map can exhibit approximately 1D
behaviour if the axis is never tangent with any  isoperiodic lines.
In table 1 we also show the periodic windows on the axis represented  by
dash-dotted line (II) in Fig. 4. It is clear that all of these
 words increase monotonically as $a$ increases except the word 10001C
 and the period windows 10111C,  1000C, 10000C and 100000C
 are missing.

\section*{\center\small\bf IV. APPLICATIONS TO ODE's}

The above idea can be extended to many other two- or higher-dimensional
systems.  Here we only take the  R\"ossler's equations$^{13}$
\begin{equation}
\begin{array}{l}
\dot x = - y - x, \\
\dot y =  x + a y, \\
\dot z = b + z ( x - c ), \\
\end{array}
\end{equation}
and the forced Brusselator$^4$
\begin{equation}
\begin{array}{l}
\dot x = A - (B + 1) x - x^2 y + \alpha \cos(\omega t), \\
\dot y =  B x - x^2 y, \\
\end{array}
\end{equation}
as examples.

The 2D attractor of the R\"ossler's equations is usually taken from a
section of the 3D flow on the half-plane $y$ = 0, $x <$ 0$^{13}$. It has
already shown$^{19}$ that the unstable  periodic orbits of the attractor
can be generated with only one kneading sequence up to period 12 for
parameters $c$ = 2, $d$ = 4 and $a$ = 0.408 (corresponding to $L \geq 12$).
 We find similar results ($L \geq 9$) for
 $c$ = 2, $d$ = 4 and $0.125 < a < 0.415$. Table 2 shows the periodic
 windows up to period 9 in descending $a$ order along with
 their periods, words and locations on the parameter axis.
 They are exactly consistent with part of the U-sequence from word C
 up to 1001011C.

\begin{table}[ht]
\caption[]{The symbolic sequences for the periodic windows
of the R\"ossler's equations}
\begin{center}
\begin{tabular}{|c|l|l|l|}
\hline
{No.} & Period & Word &  $a$ range \\
\hline
 0 &  0  & C &   0.125-0.335 \\
   &  2  & 1C &   0.336-0.375 \\
   &  2$\times$2  & 101C  &   0.376-0.3834 \\
   &  2$^2 \times$2  & 1011101C  &  0.3836-0.3852 \\
 2 &  6  & 10111C &  0.390668-0.39097 \\
 3 &  8  & 1011111C &  0.393624-0.393638 \\
 4 &  9  & 10111111C &   0.395446-0.395449 \\
 5 &  7  & 101111C &   0.396638-0.396676 \\
 6 &  9  & 10111101C &   0.398034-0.398041 \\
 7 &  5  & 1011C &  0.399948-0.399966 \\
 8 &  9  & 10110101C &  0.402190-0.402194 \\
 9 &  7  & 101101C &  0.403530-0.403564 \\
 01 &  9  & 10110111C &  0.404690-0.404691 \\
 00 &  8  & 1011011C &  0.406054 \\
 12 &  3  & 10C &  0.40912-0.41091 \\
    &  2$\times$3  & 10010C &  0.41092-0.41175 \\
 13 &  8  & 1001011C &  0.414432 \\
\hline
\end{tabular}
\end{center}
\end{table}

The forced Brusselator had been extensively studied with symbolic dynamics
of 1D maps$^{4}$. An incomplete U-sequence up to period six along the axis
 $A = 0.46 - 0.2\omega$ had already been found by Hao et. al$^{4}$
  which is also listed in Table~1. Only the periodic window 10001C was
  missing. Our investigation on the Poincar\'e map with symbolic dynamics
shows that $L=2$ for the parameter range $0.8056<\omega<0.8194$ so that
the U-sequence up to period 6 might be incomplete. Recently, J.X. Liu has
confirmed that the missing period 10001C is pruned$^{22}$.

 \section*{\center\small\bf V. CONCLUSION AND DISCUSSION}

In this paper the universal bifurcation property and its limitations of
the H\'enon map in 2D parameter space $(a,b)$ is discussed with symbolic
dynamics.  Two topological quantities $\delta$ and $L$ are defined to
characterize this topological universality. In the universal-$L$
region, as that of 1D unimodal map, there is a perfect MSS-sequence up to
 period$\leq$L along any axis  provided that the axis is never tangent
with any isoperiodic lines, though the phase portraits of the H\'enon map
exhibit very complicated  2D behaviour. Extending this idea to many other
two- or higher-dimensional systems ensures that the symbolic dynamics of
1D mappings is an effective technique to investigate  the universality in
these two- or higher-dimensional systems and then the parameter for
definite periodic motion may be predicted$^{23}$. We have presented two
examples of ordinary differential equations (ODE's), the R\"ossler
equations$^{13}$ and the forced Brusselator$^4$, to demonstrate the
validity and robustness of our approach.

It should be noted that only the short period is considered
although the theory presented in this paper is also valid for
higher period. In fact, in real experiment (or numerical study on ODE's
or PDE's), only short periodic orbit can be obtained. Our investigation
shows that it is not a surprizing result that universal sequences with
short period are found in many experiments. Moreover, our result shows
that it is also a universal phenomenon that universal sequences with
short period can be found in many nonlinear dissipative systems.  This
observation ensures that the parameter of many periodic motion for many
 dynamical systems (such as some fluid system, e. g. ref. 23) can be well
predicted.

In this paper we also show that  even on a axis out of the universal region,
the H\'enon map can exhibit approximately 1D behaviour. This observation
interprete the numerical results that in some nonlinear dynamical
systems only incomplete U-sequences had been found$^4$. Anyway, our
defined universal-M region gives the background to interprete
the experimental and numerical observations that complete or imcomplete
U-sequences with short period can be found in many dissipative systems,
and understand the limitations that 1D symbolic dynamics can be used to
study two- or high-dimensional dissipative systems.

$Acknowledgements$. The author sincerely  thanks Prof. Hao Bai-lin for his
 encouragement. He also thanks Liu Jiuxing for helpful discussion on the
forced Brusselator. This work was partially supported by a grant from CNSF.
\section*{}
{\bf References}
\begin{description}
\item[{*}]{ Present Address; email:zlin@fudan.ihep.ac.cn}.
\item[{[1]}]{ N. Metropolis, M. Stein and P. Stein,
J. Combinat. Theor. A\ {\bf 16}, 25 (1973)}.
\item[{[2]}]{ A. Libchaber, S. Fauve and C. Laroche,
 Physica D{\bf 7}, 73 (1983).}
\item[{[3]}]{ K. Coffman, W.D. McCormick and H. L. Swinney,
 Phys. Rev. Lett. {\bf 56}, 999 (1986);
 R. H. Simoyi, A. Wolf and H. L. Swinney, {\it ibid} {\bf 49}, 245 (1982).}
\item[{[4] }]{B. L. Hao,\
 Physica A \ {\bf 104}, 85 (1986)}.
\item[{[5]}]{ K. Tomita and I. Tsuda, Prog. Theor. Phys. Suppl. 69,
(1980)185.}
\item[{[6]}]{  D. R. Moore and N. O. Weiss,
  Phil. Trans. R. Soc. Lond. A\ {\bf 332}, 121 (1990);
 M. R. E. Proctor and N. O. Weiss,
  Nonlinearity\ {\bf 3}, 619 (1990)}.
\item[{[7]}]{G. H. Gunaratne, M. H. Jensen and I. Procaccia,
 Nonlinearity  {\bf 1}, 157 (1988)}.
\item[{[8]}]{G. H. Gunaratne, P. S. Linsay, M. J. Vinson,
 Phys. Rev. Lett. {\bf 63}, 0 (1989).}
\item[{[9]}]{ H.P. Fang and B.L. Hao, Symbolic Dynamics of the Lorenz
Equations.  Chaos, soliton $\&$ Fractal, accepted (1995). }
\item[{[10]}]{ M. H\'enon,\
 Commun. Math. Phys. \ {\bf 50}, 65 (1976)}.
\item[{[11]}]{C. Mira, in {\it Dynamical systems and chaos}, ed. L. Garrido,
 Lecture Notes in Phys. {\bf 179}, Springer-Verlag (1983), and the
references therein.}
\item[{[12]}]{ Jason A. C. Gallas,
 Phys. Rev.  Lett. {\bf 70}, 2714 (1993).}
\item[{[13]}]{O. E. R\"ossler's, \  Phys. Lett. A {\bf 57}, 397 (1976)}.
\item[{[14]}]{Hao Bai-lin, {\it Elementary Symbolic Dynamics and Chaos
 in Dissipative Systems},  World Scientific (1989)}.
\item[{[15]}]{ J. Milnor, W. Thurston ,
 Lect. Notes. Math. {\bf 1342}, 463 (1988)}
\item[{[16]}]{ P. Grassberger, H. Kantz,
 Phys. Lett. A {\bf 113}, 235 (1985);
 Physica D {\bf 18}, 75 (1985).}
\item[{[17]}]{ P. Cvitanovic, G. H. Gunaratne and I. Procaccia,
 Phys. Rev.  A {\bf 38}, 1503 (1988).}
 \item[{[18]}]{ G D'Alessandro, P. Grassberger, S Isola and A Politi,
 J. Phys. A {\bf 23}, 5295 (1990).}
\item[{[19]}]{H. P. Fang,  Phys. Rev. E {\bf 49}, 5025 (1994)}.
\item[{[20]}]{ L. Flepp, R. Holzner, E. Brun, M. Finardi and R. Badii,
Phys. Rev. Lett. {\bf 67}, 2244 (1991);
M. Finardi, L. Flepp, J. Parisi, R. Holzner, R. Badii and  E. Brun,
{\it ibid} \ {\bf 68}, 2989 (1992).}
\item[{[21]}]{Though we can calculate the $L$ value for given parameters,
the distribution of $L$ values are rather irregular so that we do not give
 iso-L lines. A detail discussion will be presented elsewhere.}
\item[{[22]}]{ J.X. Liu, ongoing work, unpublished.}
\item[{[23]}]{ H. P. Fang and Z. H. Liu,  Phys. Rev. E {\bf 50},
2790(1994)}.

\end{description}
\section*{}
{\bf Figure Captions}
\begin{description}
\item[]{ Fig.~1\ The Logistic map $x_{n+1} = 1 - \mu x_n^2$ with
$\mu$ = 1.75488 exhibits a 3-cycle of the type 10C.}
\item[]{ Fig. 2 \  The strange attractor  of the H\'enon map for ($a, b$)=
        (1.4, 0.16). The heavy lines outline the strange attractor. The
diamonds are the ``primary" tangencies. The dotted line connecting them
divides the full attractor into two subsets marked by 0 and 1.}
\item[]{ Fig. 3 \ The symbolic plane of the H\'enon map for ($a, b$)=
(1.4, 0.16).  (b) an enlarged part of the symbolic plane.}
\item[]{ Fig. 4 \ The isoperiodic lines together with the universal-7
region (the region between two heavy lines) and the borders for the H\'enon
map exhibit an attracting set with initial points $(x_0, y_0)$ very close
to original point $(0, 0)$ (the two heavy short dashed lines).
 The long dashed (I) and dash-dotted (II) lines represent two axes in and
out of the universal-7 region, on which a complete and incomplete
U-sequence is found respectively. (b) an  enlarged part.}
\end{description}
\end{document}